\documentclass[prd, amsfonts, onecolumn, nofootinbib, showpacs]{revtex4-2}
\usepackage{graphicx, epsfig}
\usepackage{color}
\usepackage{amsmath}
\usepackage{amssymb}
\usepackage{physics}
\usepackage[normalem]{ulem}
\newcommand{\be}{\begin{equation}}
\newcommand{\ee}{\end{equation}}
\newcommand{\bea}{\begin{eqnarray}}
\newcommand{\eea}{\end{eqnarray}}

\newcommand{\gapp}{\mathrel{\raise.3ex\hbox{$>$}\mkern-14mu
\lower0.6ex\hbox{$\sim$}}}
\newcommand{\lapp}{\mathrel{\raise.3ex\hbox{$<$}\mkern-14mu
\lower0.6ex\hbox{$\sim$}}}
\def\bbox{{\,\lower0.9pt\vbox{\hrule \hbox{\vrule height 0.2 cm
\hskip 0.2 cm \vrule  height 0.2 cm}\hrule}\,}}

\begin{document}
\title{Milky Way and M31 rotation curves:  $\Lambda$CDM vs. MOND}
\author{De-Chang Dai$^{1,3}$\footnote{communicating author: De-Chang Dai,\\ email: diedachung@gmail.com\label{fnlabel}}, Glenn Starkman$^3$, Dejan Stojkovic$^2$}
\affiliation{$^1$ Center for Gravity and Cosmology, School of Physics Science and Technology, Yangzhou University, 180 Siwangting Road, Yangzhou City, Jiangsu Province, P.R. China 225002}
\affiliation{ $^2$ HEPCOS, Department of Physics, SUNY at Buffalo, Buffalo, NY 14260-1500}
\affiliation{ $^3$  CERCA/Department of Physics/ISO, Case Western Reserve University, Cleveland OH 44106-7079}


\begin{abstract}
\widetext

{
We analyze the existing rotation-curve data of the Milky Way and M31 galaxies that extends to very large distances and low accelerations. We find a systematic downward trend in the weak acceleration (large distances) segment  of the radial acceleration. 
A similar downward trend has been noticed in the $\Lambda$CDM EAGLE simulation \cite{Dai:2017unr}, while the deviation from the generic MOND prediction would need to be ascribed to an external field effect, or possibly a {\it post facto} selected acceleration function $\mu(x)$.}
\end{abstract}


\pacs{}
\maketitle

Galaxy rotation curves were the first widely accepted evidence for the existence of dark matter \cite{Rubin:1970zza}.  Outside the maximum radius where the luminous matter contributes significantly to the total mass of a galaxy, Newtonian gravity (and General Relativity (GR)) predict that the centripetal acceleration binding tracers in orbit around the galaxy, and hence the rotational velocity, should fall, yet consistently among all galaxies it plateaus \cite{Rubin:1970zza,1980ApJ...238..471R,1981AJ.....86.1791B,1988MNRAS.234..131P,Persic:1995ru,Navarro:1996gj,Corbelli:1999af}.   The dark matter hypothesis interprets this plateau as evidence for the presence of additional non-luminous ``dark'' matter.  A number of lines of argument suggest that most, if not all, of this dark matter would need to be non-relativistic -- what cosmologists refer to as cold dark matter (CDM) \cite{Peebles:1982ff}. CDM is at the core of the concordance $\Lambda$CDM model of cosmology, and the success of $\Lambda$CDM in predicting the details of the power spectrum of temperature and polarization fluctuations of the cosmic microwave background radiation \cite{Planck:2018nkj} can be regarded as a stringent test that the dark matter hypothesis has passed.  Meanwhile, considerable theoretical, experimental and observational effort has been invested in the search for the nature of this dark matter through non-gravitational signals and distinctive gravitational signals \cite{PerezdelosHeros:2020qyt}.   

A competing hypothesis for the plateau in galaxy rotation curves, Modified Newtonian Dynamics (MOND) \cite{1983ApJ...270..365M}, is that Newton's law (and hence General Relativity) is incorrect at small accelerations.  Specific implementations of the MOND paradigm at the level of phenomenological alternatives to Newton's law or of Poisson's law have long been considered.  While these phenomenological alternatives are limited in their applicability (weak field, non-relativistic, isolated systems, often requiring spherical or axial symmetry), for those galaxies to which they can be applied they fit the rotation curves with only a single parameter per system, the mass-to-light ratio of the luminous matter, and a single universal function to match standard Newtonian behavior at large acceleration to the desired behavior at low acceleration \cite{Sanders:2002pf}.
However, the applicability of MOND to generic systems, including cosmology, has been hampered by the absence of a suitable Relativistic MOND (RMOND) theory -- a modification of GR that reduces to MOND in appropriate limits and is consistent with key cosmological observables and other tests of GR.  Recent work on such Relativistic MOND (RMOND) theories  \cite{Bekenstein:2004ne,Skordis:2020eui} may have removed that difficulty.

The discovery \cite{McGaugh:2016leg} of the Radial Acceleration Relation (RAR) -- a tight correlation between the observed acceleration of low-acceleration tracers in galaxies and the acceleration expected just from the baryonic matter -- was expected in the context of MOND.
On the other hand it is widely regarded as a mystery in $\Lambda$CDM, and therefore a failure of that theory.
However, \cite{Navarro:2016bfs,Ludlow:2016qzh} and one of the authors \cite{Dai:2017unr} independently examined data from the EAGLE simulation \cite{Crain:2015poa,McAlpine:2015tma,Schaye:2014tpa}, one of the largest cosmological hydrodynamical simulations, and found that it unexpectedly demonstrated such a tight correlation. 
Its origin must lie in physics emergent from the complex interactions of baryonic matter and the non-linear structures it forms (e.g., stars), which are not directly captured by the standard particle interactions of an N-body simulation, but which the EAGLE Project attempts to model phenomenologically.  
Unsurprisingly, the EAGLE correlation does not precisely match the observed RAR -- the underlying theory is too complicated to expect an accurate prediction from a phenomenological model that was not calibrated to it -- however this should probably be regarded as a very unexpected success of $\Lambda$CDM,  though perhaps a less compelling one than for MOND.

While the dark matter and MOND hypotheses may agree, if only by construction, on galaxy rotation curves at radii outside the luminous matter, there is no reason for them to continue to do so at larger radii where the centripetal acceleration falls much lower. 
Consider a spiral galaxy. 
Within the $\Lambda$CDM model, there are three main structures in the galaxy - a bulge, a disk and a dark halo. The innermost part is the bulge, which is the gravitationally dominant component within several $kpc$ from the center. 
The stellar disk is then dominant out to distances of several tens of kpc, where the CDM halo takes over.  
The presence of the dark matter halo is the primary reason why the rotational velocity does not decrease dramatically after the effects of ordinary matter are saturated. 
However, within $\Lambda$CDM, the effects of dark matter also eventually saturate. 
At larger radii, there is no additional component to maintain the rotational velocity at a constant value, so it should decrease outside some characteristic radius of the halo.
Within MOND, no such transition is predicted for isolated galaxies.

Most radial velocity data in galaxies cover only accelerations larger than $a_{min}\sim 10^{-11}m/s^2$, as pointed out in \cite{Navarro:2016bfs}. The region of lower accelerations has not been fully explored yet.  However, as we discuss below, 
recent data on the rotation curves of the  Milky Way (MW) and M31 \cite{2020Galax...8...37S,2012PASJ...64...75S, 2015PASJ...67...75S}  extend to larger distances and hence lower accelerations.
Intriguingly, as expected in $\Lambda$CDM, after plateauing at intermediate distances (several tens of kpc from the galactic center),
the rotation curve does not remain flat at larger radii. 
Rotation velocities decrease by tens of $km/s^2$, down to values that are much smaller than the plateau value.  
In the MW, this happens at the distances larger than several tens of kpc. 
Similar trends can also be found in the galaxy M31 (Fig.~5 in \cite{2015PASJ...67...75S}). 	

As we discuss below, this decrease in the rotation velocity at large radius is anticipated quantitatively in the results of the EAGLE simulation.  It therefore appears to be consistent with the general expectations of $\Lambda$CDM.  
On the other hand, this is not what is expected for an isolated galaxy in MOND --
there are strong indications that the rotation curves deviate from the MOND prediction for accelerations below $10^{-11}m/s^2$\cite{Dai:2017unr}. 	
Such deviations are ascribed by \cite{Chae:2020omu} to the ``external field effect'' (EFE) in MOND -- a consequence of a galaxy's environment upsetting the normal MOND expectation  (see also studies based on other modified gravity models \cite{Moffat:2014pia,Moffat:2007yg,Davari:2020ijn,Moffat:2021tfs,Mannheim:2021mhj,Davari:2021mge,Moffat:2021log}). 

Below we present rotation curves for the Milky Way and M31, and compare them to the results of the EAGLE simulation and to the predictions of MOND for isolated galaxies.

%

\section{The Milky Way and M31 as spiral galaxies in $\Lambda$CDM}
In the $\Lambda$CDM model, the rotation curve of a spiral galaxy can be well approximated as arising due to the combined contributions from and axisymmetric bulge, disk and dark halo. (Of course both real and simulated galaxies are more complicated.)
This allows us to represent the rotational velocity squared  as the sum of three components,
\begin{equation}
v^2(r)\equiv v_b^2(r)+v_d^2(r)+v_h^2(r)\,,
\end{equation} 
where $r$ is the distance from the center of the galaxy, while $v_b$, $v_d$ and $v_h$ are the contributions to $v^2$ due to the bulge, disk and dark halo respectively. 
Newtonian mechanics simply relates $v_i(r)$ (for $i=b,d,h$) to the mass of that component within the radius r,
\begin{equation}
	v_i^2=\frac{G M_i(r)}{r}\,.
\end{equation}

A galactic bulge is taken to be spherically symmetric with a de Vaucouleur profile \cite{1948AnAp...11..247D}. 
Its surface mass density is 
\begin{equation}
	\Sigma_B(r) = \Sigma_{be}e^{-\kappa \left( (\frac{r}{a_B})^{1/4}-1\right)},
\end{equation} 
where $\kappa=7.6695$, and $\Sigma_{be}$ is the surface mass density at the half-mass radius $r=a_B$. 
The mass density of the bulge is then
\begin{equation}
	\rho_b(r)=-\frac{1}{\pi} \int_r^\infty\frac{d\Sigma_B (x)}{dx} \frac{1}{\sqrt{x^2-r^2}}dx\,,
\end{equation}
so that the bulge mass within the radius $r$ is 
\begin{equation}
	M_b(r)=4\pi \int ^r_0 \rho_b(r) r^2 dr\,
\end{equation}
and
\begin{equation}
	v^2_b(r)=\frac{4\pi G}{r} \int ^r_0 \rho_b(r) r^2 dr\,.
\end{equation}

A galactic disk can be approximated by an exponential disk \cite{1959HDP....53..311D,1970ApJ...160..811F} with surface mass density 
\begin{equation}
	\Sigma_d(r)=\Sigma_0 \exp\Big (-\frac{r}{a_d}\Big)\,.
\end{equation}
$\Sigma_0$ is the central value and $a_d$ is a scale radius. 
The rotation velocity squared due to the disc \cite{1970ApJ...160..811F} can be written explicitly,
\begin{equation}
	v_d^2=
		\pi G \Sigma_0 \frac{r^2}{a_d} 
		\left[I_0\left(\frac{r}{2a_d}\right)
			K_0\left(\frac{r}{2a_d}\right)
			-I_1\left(\frac{r}{2a_d}\right)
			K_1\left(\frac{r}{2a_d}\right)
			\right]\,,
\end{equation}
where $I_i$ and $K_i$ are modified Bessel functions. 

A dark halo is taken to follow the NFW profile \cite{1996ApJ...462..563N}, 
\begin{eqnarray}
\rho_h (r)&=&\frac{\rho_0}{\frac{r}{h}
	\left(1+\frac{r^2}{h^2}\right)} \,,
\end{eqnarray}
where $\rho_0$ and $h$ are the scale density and scale radius of the dark halo respectively. 
This leads to
\begin{equation}
	M_h(r)=4\pi \rho_0 h^3 
	\left[\ln\left(1+\frac{r}{h}\right)
	-\frac{r}{r+h}\right]\,,
\end{equation}   
and
\begin{equation}
v_h^2 =\frac{G M_h(r)}{r}
\end{equation}

The total acceleration is then 
\begin{equation}
\label{a_total}
a_t =\frac{v_b^2+v_d^2+v_h^2}{r}\,,
\end{equation}
and the expected rotation curve is
\begin{equation}
\label{a_total}
v_t(r) =\sqrt{v_b^2+v_d^2+v_h^2} \,.
\end{equation}

\begin{figure}[h]
\includegraphics[width=10cm]{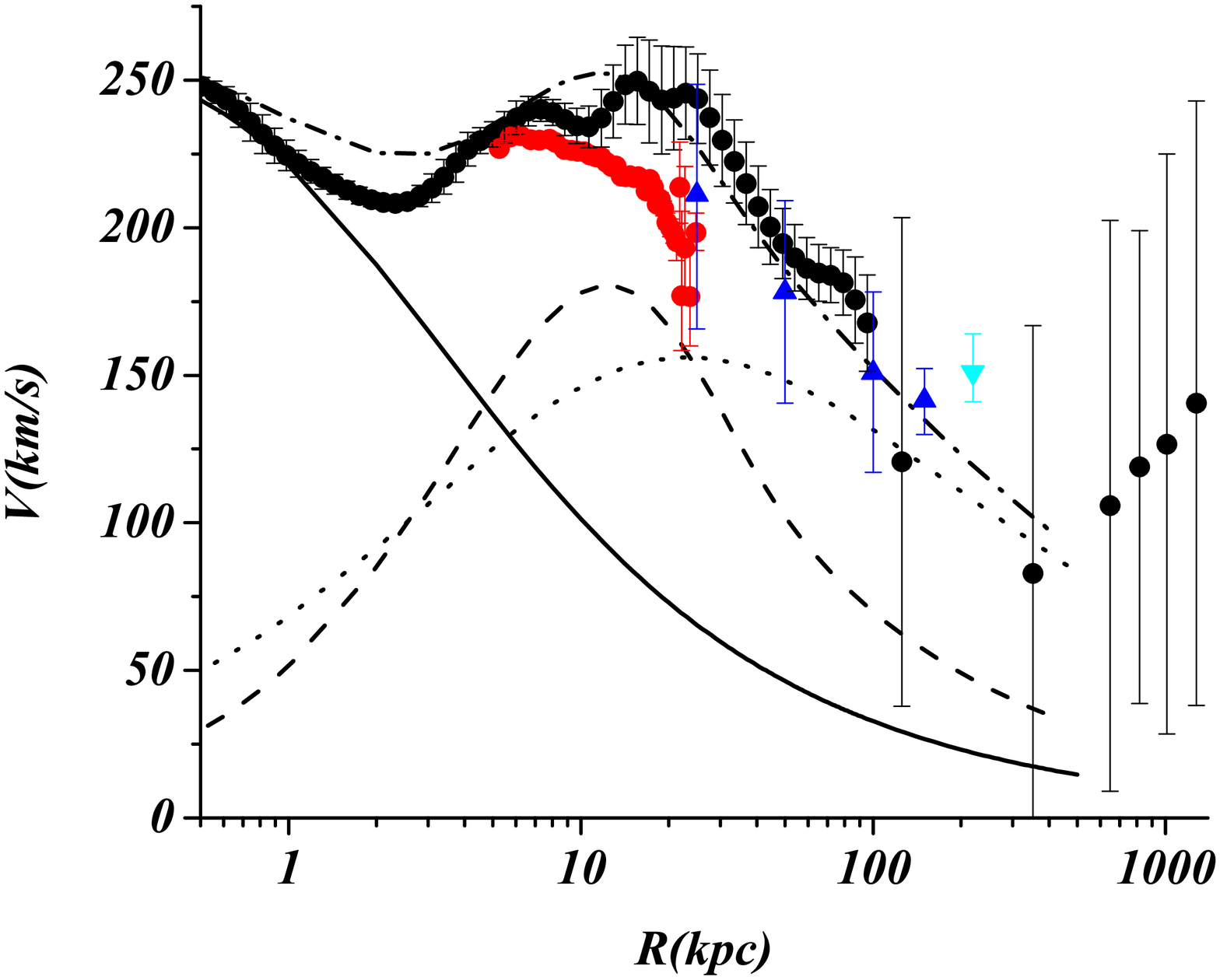}
\caption{The rotation curve of the Milky Way. The data (solid dark circles with error bars) for $r< 100$kpc come from \cite{2020Galax...8...37S}, while for $r>100$kpc  from \cite{2012PASJ...64...75S}. The solid, dashed and doted lines describe the contribution from the bulge, stellar disk and dark matter halo respectively, within a $\Lambda$CDM model of the galaxy. The dashed-dot line is the total contribution of all three components. The parameters of each component are taken from \cite{2015PASJ...67...75S}.  For comparison, the Milky way rotation curve from GAIA data releaese II is shown in color. The red dots are data from \cite{2019ApJ...871..120E}, the blue upward-pointing triangles are from \cite{2019ApJ...875..159E}, while the cyan downward-pointing triangles are from \cite{Callingham:2018vcf}.
}
\label{milkyway}
\end{figure}

\begin{figure}[h]
\includegraphics[width=10cm]{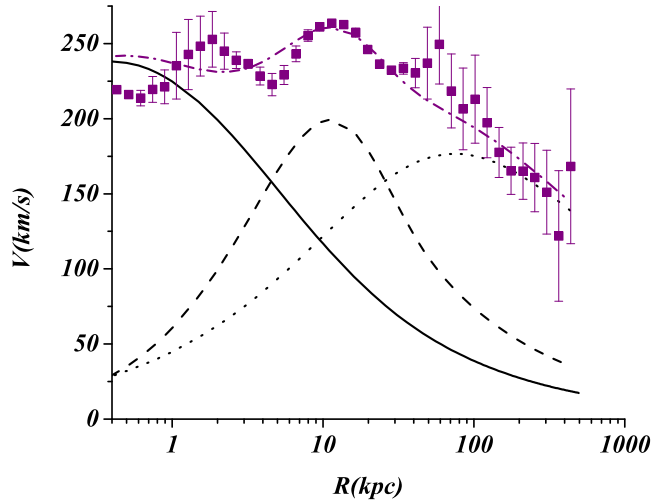}
\caption{The rotation curve of the M31. The data (solid purple squares with error bars) come from \cite{2015PASJ...67...75S}. The solid, dashed and doted lines describe the contribution from the bulge, stellar disk and dark matter halo respectively, within a $\Lambda$CDM model of the galaxy. The dashed-dot line is the total contribution of all three components. The parameters of each component are taken from \cite{2015PASJ...67...75S}. 
}
\label{M31}
\end{figure}

There are $6$ parameters ($a_B$, $a_d$, $h$, $\Sigma_{be}$, $\Sigma_0$ and $\rho_0$) that fully specify all three components of a rotationally supported galaxy in $\Lambda$CDM.
We use the values for the MW and M31 from \cite{2015PASJ...67...75S}. 
For MW they are $a_B=0.87\pm 0.07$kpc, $a_d=5.73\pm 1.23$kpc, $h=10.7\pm 2.9$kpc, $\Sigma_{be}=0.25\pm 0.02 \times 10^{11}M_\odot$, $\Sigma_0=1.12\pm 0.4 \times 10^{11}M_\odot$ and $\rho_0=18.2\pm 7.4 \times 10^{-3} M_\odot pc^{-3}$. 
For M31 they are $a_B=1.35\pm 0.02$kpc, $a_d=5.28\pm 0.25$kpc, $h=34.6\pm 2.1$kpc, $\Sigma_{be}=0.35\pm 0.004 \times 10^{11} M_\odot$, $\Sigma_0=1.26\pm 0.08\times 10^{11}M_\odot$ and $\rho_0=2.23\pm -.24 \times 10^{-3}M_\odot pc^{-3}$.
 Figure \ref{milkyway} and \ref{M31} show the rotation curve of the Milky Way and M31 respectively, and its resolution into the bulge, disk and halo components.
The data for $r< 100$kpc come from \cite{2020Galax...8...37S}, while for $r>100$kpc  from \cite{2012PASJ...64...75S}. 
As expected of a low-dimensional axisymmetric approximation of a spiral galaxy, the fit to the model reproduces the gross features of the rotation curve reasonably well, but not the fine details.  Recent Milky way rotation curve from GAIA data release II are plotted in figure \ref{milkyway} in color \cite{2019ApJ...871..120E,2019ApJ...875..159E,Callingham:2018vcf,2020MNRAS.494.4291C}. The general trend is very similar to the  rotation curve based the older data \cite{2015PASJ...67...75S}.  We see that while the GAIA rotation curve at modest galactic radii differs significantly from earlier data, so far, there is no statistically significant change at larger radii where the model suggests that dark matter halo dominates the kinematics.
One could try to model the galaxy with GAIA data in the regime where the data is available,  however the behavior at modest galactic radii will not have significant influence at larger radii where our analysis is concentrated. We also note that there is no such discrepancy in M31 galaxy data.

\section{Comparing  to $\Lambda$CDM and MOND}

To compare observations with $\Lambda$CDM and MOND predictions, the Newtonian gravitational acceleration due to the ordinary matter must be separated from the total acceleration. Presumably the acceleration due to baryons, $a_{B}$, comes from the bulge and the disk (with no dark matter halo contribution), i.e. 
\begin{equation}
	a_{B}(r) =\frac{v_b^2(r)+v_d^2(r)}{r}
\end{equation} 

The MOND-predicted acceleration is obtained from \cite{1983ApJ...270..365M}
\begin{equation}
	\label{eqn:MOND}
	\mu(a/a_0)\vec{a}=\vec{a}_{B}\,,
\end{equation}
where $a_0$ is a critical acceleration constant, which we will set to $a_0 = 1.2\times 10^{-10}m/s^2$, while $\mu$ is an empirical function which is often modeled with 
\begin{equation}
	\mu(x)=\frac{x}{1+x} \,.
\end{equation} 
MOND then predicts the total acceleration to be
\begin{equation}
	\label{eqn:aMOND}
	a_M=\frac{a_{B}+\sqrt{a_{B}^2+4a_0 a_{B}}}{2} \,,
\end{equation} 
which is to be compared with \eqref{a_total} in $\Lambda$CDM.

In fact, this prediction assumes that the galaxies being observed are isolated.
The MOND isolated galaxy (MIG) prediction would be expected to fail where tidal accelerations
due to other structures in the environment disrupt the symmetries of the system.
This external field effect (EFE) is a longstanding prediction of MOND, but its details  -- what specific tidal field modifies \eqref{eqn:aMOND} and how  -- would depend on how \eqref{eqn:MOND} emerges from a specific relativistic theory of broader applicability.

\begin{figure}[t]
\includegraphics[width=10cm]{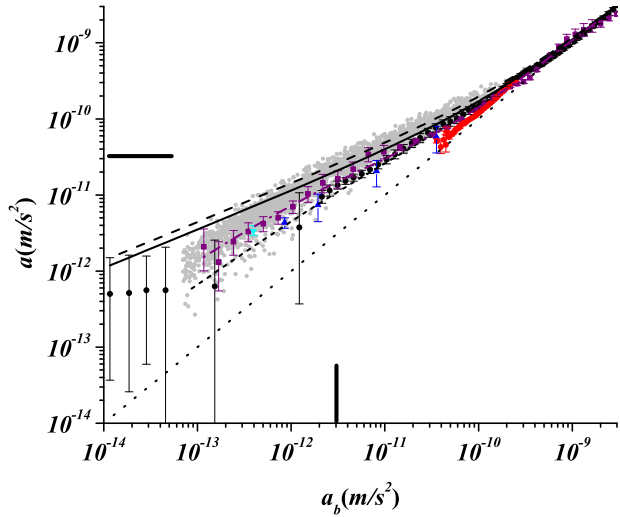}
\caption{The total acceleration, $a$ vs. the Newtonian acceleration due to baryons, $a_B$, for data and models.
The black circles with error bars represent  Milky Way data.  The purple squares with error bars represent  M31.
The $\Lambda$CDM fit to them is the short-dashed line.
(The dash-dot line is the $\Lambda$CDM fitting curve of M31.)
The dotted line is the reference line for $a=a_B$. 
The dashed and solid  lines are predicted by MOND with $a_0=1.2\times 10^{-10}m/s^2$ and $a_0=1.9\times 10^{-10}m/s^2$. 
 The gray dots are from the EAGLE simulation (data file RefL0025N0752 on EAGLE's website) of $\Lambda$CDM in \cite{Dai:2017unr}. 
 The thick horizontal (vertical) line crossing the a ($a_B$) axis marks the acceleration (baryonic acceleration) below which very little data is available -- other than for the MW and M31. Accelerations in M31 and the MW are smaller than the MOND prediction at lower values of $a_B$. 
A potentially observable discrepancy appears at the radius of several tens of kpc for both galaxies.  We plot the results out to $500 kpc$.  The galaxy mass in EAGLE's data is chosen to be between $5\times 10^{ 10} M_\odot $ to $5\times 10^{11} M_\odot $. For comparison, the Milky way rotation curve from GAIA data release II are shown in color. The red dots are data from \cite{2019ApJ...871..120E}, the blue triangles are from \cite{2019ApJ...875..159E}, while the cyan down triangles are from \cite{Callingham:2018vcf}.
}
\label{milkyway_MOND}
\end{figure}

In $\Lambda$CDM, disks of  galaxies usually form at the centers of dark matter halos. They span a relatively narrow range of acceleration, and the acceleration profiles are self-similar from $a_0\sim 10^{-10} m/s^2$ to $a_{min}\sim 10^{-11}m/s^2$ \cite{Navarro:2016bfs}. 
$\Lambda$CDM predictions for the rotation curve then clearly deviate from MOND predictions only for $a\lesssim a_{min}$\cite{Dai:2017unr}. Because this is within the dark-matter-dominated region, it is not easily accessible to observations. Since there are very few distant galaxies in which we can explore  accelerations below $a_{min}$, this deviation frequently goes unnoticed. However, in nearby galaxies like our own Milky Way and M31, we can can explore such low accelerations  and the distinguishable predictions of MOND and $\Lambda$CDM, as shown in Fig.~\ref{milkyway_MOND}. 

Fig.~\ref{milkyway_MOND} shows that for smaller radii, i.e. larger accelerations, MOND does not significantly deviate from expected $\Lambda$CDM model galaxies, but that this changes for lower accelerations where the difference is clear.  This feature may not be easy to observe for galaxies that are further away, because the discrepancy becomes clear only in dark halo regions far away from the bulk of the visible matter.    It should be noted that the model rotation curve of M31 is closer to the MOND predictions than that of the Milky May at lower accelerations. This is because M31's halo is more spread out. ($h=34.6 kpc$ for M31 compared to $h=10.7 kpc$ for the  Milky Way.) 
These lower accelerations are also where tidal fields are more likely to induce an external field effect in the MOND theory.

\begin{figure}[t]
\includegraphics[width=10cm]{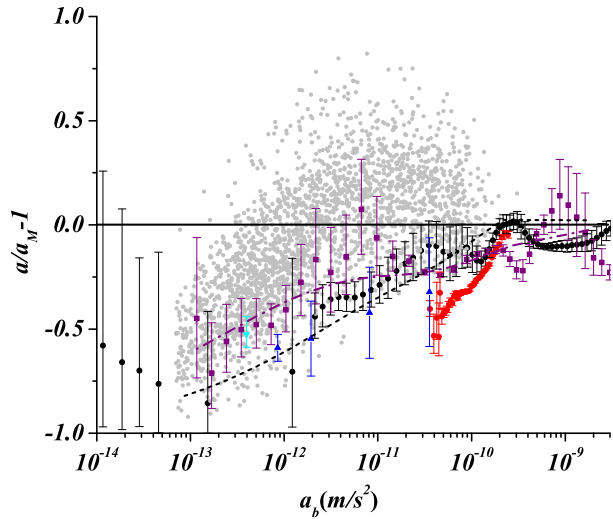}
\caption{Figure \ref{milkyway_MOND} recast as a comparison between the total acceleration, $a$, and the MOND prediction, $a_M$, as a function of the acceleration due to baryons $a_B$. The solid horizontal line is $a=a_M$.   The circles and squares with error bars represent the Milky Way and M31 data, while the gray dots are from the EAGLE simulation of $\Lambda$CDM in \cite{Dai:2017unr}. For $a_B\gtrsim10^{-10}m/s^2$ any difference between $a$ and $a_M$is unclear. However, once $a_B$ drops well below $10^{-11}m/s^2$, the discrepancy emerges.  The short-dashed line is the $\Lambda$CDM fitting curve of the MW.  The dash-dot line is the $\Lambda$CDM fitting curve of M31.  The mass range of galaxies in EAGLE's data is chosen to be between $5\times 10^{ 10} M_\odot $ to $5\times 10^{11} M_\odot $. For comparison, the Milky way rotation curve from GAIA data release II is shown in color. The red dots are data from \cite{2019ApJ...871..120E}, the blue triangles are from \cite{2019ApJ...875..159E}, while the cyan down triangles are from \cite{Callingham:2018vcf}. { While the EAGLE simulation does not match the data perfectly, these plots indicate that it is much easier to accommodate a systematic downward trend with the $\Lambda$CDM model than with MOND.}
}
\label{difference}
\end{figure}

Fig. \ref{milkyway_MOND}  also compares the total observed acceleration $a$ with the acceleration $a_B$ caused by the baryons for the case of the  MW (for which the low-acceleration data is much less noisy than for M31). 
The observational data (black squares with error bars) follow the MOND isolated galaxies (MIG) prediction (dashed and solid lines) very well for larger values of acceleration (down to a few $10^{-10}m/s^2$);
however, for smaller accelerations, the observed acceleration is lower than the MIG prediction. 
Fig. 1 of \cite{Chae:2020omu} does not extend to the same low accelerations as does this plot with Milky Way data, which might be the reason why the discrepancy is not striking in \cite{Chae:2020omu}. 
Fig. \ref{difference} explicitly shows this difference by plotting $a/a_M-1$:
MOND predicts accelerations that are much larger than  observed, and the theory and data appear well-separated for $a<10^{-11}m/s^2$. 
This is the effect that was found in data in \cite{Chae:2020omu} and ascribed to the expected external field effect (EFE) -- the fact that tidal fields due to nearby mass concentrations (e.g. other galaxies) disrupt the MIG prediciton \eqref{eqn:aMOND}.

In contrast,  the MW and M31 models, like the MW data, are all alike in falling below the MOND prediction at low $a_B$.
One might complain that it is not surprising that the MW model match the data, however
Fig. \ref{milkyway_MOND} also includes the data from EAGLE galaxies \cite{Crain:2015poa,McAlpine:2015tma,Schaye:2014tpa} as presented in \cite{Dai:2017unr}.
The turndown in $a$ at low $a_B$ matches the feature previously pointed out by one of us \cite{Dai:2017unr} in the  EAGLE simulation \cite{Crain:2015poa,McAlpine:2015tma,Schaye:2014tpa}.
We see that for EAGLE $a$ vs. $a_B$ follows a trend similar to that in MOND for $a\gtrsim a_{min}$ with reasonably low dispersion, but then $a$ falls faster at low $a_B$ in EAGLE than predicted by MOND, and the dispersion in the simulated data increases.
This broad feature is consistent with the data, and with the fits to simple dark-matter models.

 We note here that the EAGLE simulation data points in figure Fig. \ref{difference} are a combination of many independent galaxies. The rotation curves of galaxies with less dark matter are closer to Newtonian physics. On the other hand, the rotation curves of galaxies with more dark matter are well above the MOND prediction. In general, MOND predicts some average behavior, instead of a precise behavior for every single galaxy. This is the reason why one can see single-galaxy deviations from MOND clearly, while the deviation is not so clear if many galaxies are included.

It would be attractive to be able to compare the data to the EAGLE simulation quantitatively at these low $a$ and $a_B$;
however this is not necessarily justified, and  we certainly resist doing so here.
The baryonic physics that, in EAGLE, results in the MOND-like radial acceleration relation and in the downturn and increased (fractional) dispersion at low $a_B$,  is 
microphysics that is modeled phenomenologically.
Even if we were convinced that all and only the correct relevant physics has been captured in EAGLE, 
the appropriate Bayesian hierarchal model comparison needed to draw a statistically robust conclusion would certainly require
more sophisticated simulation or emulation infrastructure than is currently extant.
Nevertheless, we think the case is clear that where ultra-low-acceleration galaxy-rotation-curve data exist they display both the RAR-like correlation of $a$ and $a_B$ expected from EAGLE at moderately low accelerations, and the stronger tail off expected at the lowest accelerations.

 Similar indications come from other studies. The authors of \cite{2017ApJ...835L..17K} found that $\Lambda$CDM  reproduces the acceleration relation predicted by MOND, and conjectured that this relation was a consequence of dissipative collapse of  baryons. Such baryonic physics is not directly obtained by individual baryonic particles in the simulations colliding and emitting photons into the simulation volume, rather it is modeled phenomenologically. 
However, their study extended only to accelerations  $\gtrsim 10^{-11.5}$ m/s$^2$, where we do not expect to see a clear deviation from RAR. In \cite{2019ApJ...882...46W} the authors found that the deviation from RAR becomes clear for radii larger than 100kpc, which agrees with our findings. 

 Meanwhile, the RAR has been studied using weak-lensing data for released by the Kilo-Degree Survey, first \cite{Brouwer:2016dvq} using KiDS GGL on isolated foreground galaxies from 180 deg2 of the Galaxy and Mass Assembly (GAMA) survey, and more recently \cite{Brouwer:2021nsr} using a selection of 1 million foreground galaxies from KiDS-1000 to achieve a fivefold increase in survey area.   The GAMA survey results were broadly consistent with the MOND prediction for an extension of the RAR down to $a\simeq10^{-12.5}$ m/s$^2$ with the observed stellar baryons plus inferred cold gas;  the KiDS-1000 yielded $a/a_b$ that were systematically higher, matching the MOND prediction with the additional inclusion of a hot gas component.  All of them are well above the MW and M31 RARs at $10^{-12.5}-10^{-11}$m/s$^2$.  Only the mean RAR  is reported. These results appear to be in tension with our analysis here.
 However, as the authors of \cite{Brouwer:2016dvq} state, a fundamental limitation of their analysis is that the additional diffuse gas surrounding galaxies is difficult to measure, and has not been included in their study. The existence of gaseous halos not included in their study would enhance acceleration from baryons (and thus $a_M$), and the measured rotation curve will be well below what MOND predicts.


In summary, flat galactic rotation curves can be explained either by introducing a dark matter component, like in the $\Lambda$CDM model, or by modifying Newtonian dynamics, like in MOND theories. However, it has been noticed that the acceleration systematically goes down at large distances, i.e. low accelerations.  This systematic downward trend in the weak gravity segment  of the galactic rotation curves was interpreted in \cite{Chae:2020omu} as an external field effect within the framework of the MOND theories. However, this feature was observed in the $\Lambda$CDM model without invoking any modification of Newtonian gravity, as was first found by one of the authors in the EAGLE simulation \cite{Dai:2017unr}. They showed that dark halos in $\Lambda$CDM generate an acceleration feature like MOND predicts, but that in the even-larger-radius, even-lower-acceleration regions, the dark halos are saturated and the acceleration then decreases as Kepler's  Laws, i.e. Newtonian gravity, but not MOND predict. We corroborated this finding here by analyzing the Milky Way and M31 data.  At very large distances, these data are easier to accommodate in the $\Lambda$CDM model. 

\begin{acknowledgments}
 D.C Dai is supported by the National Natural Science Foundation of China  (Grant No. 11775140). D.S. is partially supported by the US National Science Foundation, under Grant No.  PHY-2014021.  GDS is supported by a Department of Energy grant DE-SC0009946 to the particle astrophysics theory group at CWRU. 
\end{acknowledgments}


\begin{thebibliography}{99}

\bibitem{Dai:2017unr}
D.~C.~Dai and C.~Lu,
Phys. Rev. D \textbf{96}, no.12, 124016 (2017)
doi:10.1103/PhysRevD.96.124016
[arXiv:1712.01654 [gr-qc]].



\bibitem{Chae:2020omu}
K.~H.~Chae, F.~Lelli, H.~Desmond, S.~S.~McGaugh, P.~Li and J.~M.~Schombert,
Astrophys. J. \textbf{904}, no.1, 51 (2020)
[erratum: Astrophys. J. \textbf{910}, no.1, 81 (2021)]
doi:10.3847/1538-4357/abbb96
[arXiv:2009.11525 [astro-ph.GA]].


\bibitem{Rubin:1970zza}
V.~C.~Rubin and W.~K.~Ford, Jr.,
Astrophys. J. \textbf{159}, 379-403 (1970)
doi:10.1086/150317



\bibitem[Rubin et al.(1980)]{1980ApJ...238..471R} V.~C. Rubin, W.~K. Ford, \& N. Thonnard, apj, 238, 471. (1980) doi:10.1086/158003 
\bibitem[Bosma(1981)]{1981AJ.....86.1791B}  A. Bosma, aj, 86, 1791 (1981). doi:10.1086/113062

\bibitem[Persic \& Salucci(1988)]{1988MNRAS.234..131P}  M. Persic \& P. Salucci, Mon. Not. Roy. Astron. Soc. \textbf{234}, 131 (1988). doi:10.1093/mnras/234.1.131


\bibitem{Persic:1995ru}
M.~Persic, P.~Salucci and F.~Stel,
Mon. Not. Roy. Astron. Soc. \textbf{281}, 27 (1996)
doi:10.1093/mnras/278.1.27
[arXiv:astro-ph/9506004 [astro-ph]].

\bibitem{Navarro:1996gj}
J.~F.~Navarro, C.~S.~Frenk and S.~D.~M.~White,
Astrophys. J. \textbf{490}, 493-508 (1997)
doi:10.1086/304888
[arXiv:astro-ph/9611107 [astro-ph]].

\bibitem{Corbelli:1999af}
E.~Corbelli and P.~Salucci,
Mon. Not. Roy. Astron. Soc. \textbf{311}, 441-447 (2000)
doi:10.1046/j.1365-8711.2000.03075.x
[arXiv:astro-ph/9909252 [astro-ph]].




\bibitem{Peebles:1982ff}
P.~J.~E.~Peebles,
Astrophys. J. Lett. \textbf{263}, L1-L5 (1982)
doi:10.1086/183911

\bibitem{Planck:2018nkj}
N.~Aghanim \textit{et al.} [Planck],
Astron. Astrophys. \textbf{641}, A1 (2020)
doi:10.1051/0004-6361/201833880
[arXiv:1807.06205 [astro-ph.CO]].

\bibitem{PerezdelosHeros:2020qyt}
C.~P\'erez de los Heros,
Symmetry \textbf{12}, no.10, 1648 (2020)
doi:10.3390/sym12101648
[arXiv:2008.11561 [astro-ph.HE]].


\bibitem[Milgrom(1983)]{1983ApJ...270..365M} Milgrom, M.\ 1983, \apj, 270, 365

\bibitem{Sanders:2002pf}
R.~H.~Sanders and S.~S.~McGaugh,
Ann. Rev. Astron. Astrophys. \textbf{40}, 263-317 (2002)
doi:10.1146/annurev.astro.40.060401.093923
[arXiv:astro-ph/0204521 [astro-ph]].

\bibitem{Bekenstein:2004ne}
J.~D.~Bekenstein,
Phys. Rev. D \textbf{70}, 083509 (2004)
[erratum: Phys. Rev. D \textbf{71}, 069901 (2005)]
doi:10.1103/PhysRevD.70.083509
[arXiv:astro-ph/0403694 [astro-ph]].

\bibitem{Skordis:2020eui}
C.~Skordis and T.~Zlosnik,
[arXiv:2007.00082 [astro-ph.CO]].

\bibitem{McGaugh:2016leg} S.~McGaugh, F.~Lelli and J.~Schombert, Phys. Rev. Lett. \textbf{117}, no.20, 201101 (2016)


\bibitem{Navarro:2016bfs}
J.~F.~Navarro, A.~Ben\'\i{}tez-Llambay, A.~Fattahi, C.~S.~Frenk, A.~D.~Ludlow, K.~A.~Oman, M.~Schaller and T.~Theuns,
Mon. Not. Roy. Astron. Soc. \textbf{471}, no.2, 1841-1848 (2017)
doi:10.1093/mnras/stx1705
[arXiv:1612.06329 [astro-ph.GA]].

\bibitem{Crain:2015poa}
R.~A.~Crain, J.~Schaye, R.~G.~Bower, M.~Furlong, M.~Schaller, T.~Theuns, C.~D.~Vecchia, C.~S.~Frenk, I.~G.~McCarthy and J.~C.~Helly, \textit{et al.}
Mon. Not. Roy. Astron. Soc. \textbf{450}, no.2, 1937-1961 (2015)
doi:10.1093/mnras/stv725
[arXiv:1501.01311 [astro-ph.GA]].

\bibitem{McAlpine:2015tma}
S.~McAlpine, J.~C.~Helly, M.~Schaller, J.~W.~Trayford, Y.~Qu, M.~Furlong, R.~G.~Bower, R.~A.~Crain, J.~Schaye and T.~Theuns, \textit{et al.}
Astron. Comput. \textbf{15}, 72-89 (2016)
doi:10.1016/j.ascom.2016.02.004
[arXiv:1510.01320 [astro-ph.GA]].

\bibitem{Schaye:2014tpa}
J.~Schaye, R.~A.~Crain, R.~G.~Bower, M.~Furlong, M.~Schaller, T.~Theuns, C.~Dalla Vecchia, C.~S.~Frenk, I.~G.~McCarthy and J.~C.~Helly, \textit{et al.}
Mon. Not. Roy. Astron. Soc. \textbf{446}, 521-554 (2015)
doi:10.1093/mnras/stu2058
[arXiv:1407.7040 [astro-ph.GA]].


\bibitem[Sofue(2020)]{2020Galax...8...37S} Sofue, Y.\ 2020, Galaxies, 8, 37
\bibitem[Sofue(2012)]{2012PASJ...64...75S} Sofue, Y.\ 2012, pasj, 64, 75
\bibitem[Sofue(2015)]{2015PASJ...67...75S} Sofue, Y.\ 2015, pasj, 67, 75

\bibitem[de Vaucouleurs(1948)]{1948AnAp...11..247D}G. de Vaucouleurs, Annales d'Astrophysique, 11, 247 (1948)

\bibitem[de Vaucouleurs(1959)]{1959HDP....53..311D}  G. de Vaucouleurs, Handbuch der Physik, 53, 311 (1959). doi:10.1007/978-3-642-45932-0\_8
\bibitem[Freeman(1970)]{1970ApJ...160..811F}K.~C. Freeman, , apj, 160, 811 (1970). doi:10.1086/150474

\bibitem[Navarro et al.(1996)]{1996ApJ...462..563N} Navarro, J.~F., Frenk, C.~S., \& White, S.~D.~M.\ 1996, apj, 462, 563


\bibitem{Ludlow:2016qzh}
A.~D.~Ludlow, A.~Benitez-Llambay, M.~Schaller, T.~Theuns, C.~S.~Frenk, R.~Bower, J.~Schaye, R.~A.~Crain, J.~F.~Navarro and A.~Fattahi, \textit{et al.}
Phys. Rev. Lett. \textbf{118}, no.16, 161103 (2017)
doi:10.1103/PhysRevLett.118.161103
[arXiv:1610.07663 [astro-ph.GA]].


\bibitem[Keller  Wadsley(2017)]{2017ApJ...835L..17K} Keller, B.~W. \& Wadsley, J.~W.\ 2017, Astrophysical Journal Letter, 835, L17. doi:10.3847/2041-8213/835/1/L17

\bibitem[Wheeler et al.(2019)]{2019ApJ...882...46W} Wheeler, C., Hopkins, P.~F., \& Dor{\'e}, O.\ 2019, \apj, 882, 46. doi:10.3847/1538-4357/ab311b



\bibitem{Brouwer:2016dvq}
M.~M.~Brouwer, M.~R.~Visser, A.~Dvornik, H.~Hoekstra, K.~Kuijken, E.~A.~Valentijn, M.~Bilicki, C.~Blake, S.~Brough and H.~Buddelmeijer, \textit{et al.}
Mon. Not. Roy. Astron. Soc. \textbf{466}, no.3, 2547-2559 (2017)
doi:10.1093/mnras/stw3192
[arXiv:1612.03034 [astro-ph.CO]].

\bibitem{Brouwer:2021nsr}
M.~M.~Brouwer, K.~A.~Oman, E.~A.~Valentijn, M.~Bilicki, C.~Heymans, H.~Hoekstra, N.~R.~Napolitano, N.~Roy, C.~Tortora and A.~H.~Wright, \textit{et al.}
Astron. Astrophys. \textbf{650}, A113 (2021)
doi:10.1051/0004-6361/202040108
[arXiv:2106.11677 [astro-ph.GA]].

\bibitem[Eilers et al.(2019)]{2019ApJ...871..120E} Eilers, A.-C., Hogg, D.~W., Rix, H.-W., et al.\ 2019, apj, 871, 120. doi:10.3847/1538-4357/aaf648

\bibitem[Eadie \& Juri{\'c}(2019)]{2019ApJ...875..159E} Eadie, G. \& Juri{\'c}, M.\ 2019, apj, 875, 159. doi:10.3847/1538-4357/ab0f97

\bibitem{Callingham:2018vcf}
T.~Callingham, M.~Cautun, A.~J.~Deason, C.~S.~Frenk, W.~Wang, F.~A.~G\'omez, R.~J.~J.~Grand, F.~Marinacci and R.~Pakmor,
doi:10.1093/mnras/stz365
[arXiv:1808.10456 [astro-ph.GA]].

\bibitem[Cautun et al.(2020)]{2020MNRAS.494.4291C} Cautun, M., Ben{\'\i}tez-Llambay, A., Deason, A.~J., et al.\ 2020, mnras, 494, 4291. doi:10.1093/mnras/staa1017


\bibitem{Moffat:2014pia}
J.~W.~Moffat and V.~T.~Toth,
Phys. Rev. D \textbf{91}, no.4, 043004 (2015)
doi:10.1103/PhysRevD.91.043004
[arXiv:1411.6701 [astro-ph.GA]].


\bibitem{Moffat:2007yg}
J.~W.~Moffat and V.~T.~Toth,
Astrophys. J. \textbf{680}, 1158 (2008)
doi:10.1086/587926
[arXiv:0708.1935 [astro-ph]].

\bibitem{Davari:2020ijn}
Z.~Davari and S.~Rahvar,
Mon. Not. Roy. Astron. Soc. \textbf{496}, no.3, 3502-3511 (2020)
doi:10.1093/mnras/staa1660
[arXiv:2006.06032 [astro-ph.GA]].

\bibitem{Moffat:2021tfs}
J.~W.~Moffat and V.~T.~Toth,
Eur. Phys. J. C \textbf{81}, no.9, 836 (2021)
doi:10.1140/epjc/s10052-021-09632-5
[arXiv:2109.11133 [gr-qc]].

\bibitem{Mannheim:2021mhj}
P.~D.~Mannheim and J.~W.~Moffat,
Int. J. Mod. Phys. D \textbf{30}, no.14, 2142009 (2021)
doi:10.1142/S0218271821420098
[arXiv:2103.13972 [gr-qc]].

\bibitem{Davari:2021mge}
Z.~Davari and S.~Rahvar,
Mon. Not. Roy. Astron. Soc. \textbf{507}, no.3, 3387-3399 (2021)
doi:10.1093/mnras/stab2350
[arXiv:2108.00266 [astro-ph.CO]].

\bibitem{Moffat:2021log}
J.~W.~Moffat and V.~Toth,
Universe \textbf{7}, no.10, 358 (2021)
doi:10.3390/universe7100358
[arXiv:2104.12806 [gr-qc]].
\end{thebibliography}
\end{document}